\documentclass{SciPost}

\usepackage{svg}
\usepackage{xcolor}
\usepackage{float}
\usepackage{comment}

 \DeclareMathOperator\tr{Tr}

\begin{document}

\begin{center}{\Large \textbf{
Thermalisation as Diffusion in Hilbert Space.
}}\end{center}

\begin{center}
A.V. Lunkin\textsuperscript{1*}
\end{center}

\begin{center}
{\bf 1} Nanocenter CENN, Jamova 39, Ljubljana, SI-1000, Slovenia
\\
* Aleksey.Lunkin@nanocenter.si
\end{center}

\begin{center}
\today
\end{center}


\section*{Abstract}
{\bf 
We develop a microscopic theory of thermalisation for a thermometer coupled to a many-body bath beyond standard Markovian and Fermi-golden-rule assumptions. By modeling interaction matrix elements in the non-interacting basis as independent random variables, we derive a diffusion-propagator expression for the reduced dynamics and show that relaxation is controlled by the distribution of interaction-induced level broadenings. The theory predicts a thermalisation timescale set by the inverse typical broadening and yields a non-Markovian generalization of global balance. Exact-diagonalization tests for heavy-tailed L{\'e}vy couplings, an all-to-all transverse-field Ising model, and the one-dimensional Imbrie model show good agreement with these predictions.
}

\vspace{10pt}
\noindent\rule{\textwidth}{1pt}
\tableofcontents\thispagestyle{fancy}
\noindent\rule{\textwidth}{1pt}
\vspace{10pt}

\section{Introduction}
\label{sec:introduction}

Understanding thermalisation in closed quantum systems remains a central problem in modern physics. A common starting point is the eigenstate thermalisation hypothesis (ETH) \cite{deutsch1991quantum, srednicki1994chaos, srednicki1999approach}, which models matrix elements of local observables in the many-body eigenbasis as random variables whose variance varies smoothly with energy. For reviews, see Refs.~\cite{d2016quantum, deutsch2018eigenstate}. ETH explains why long-time local observables approach thermal values for states with narrow energy support. However, ETH alone does not provide controlled predictions for thermalisation times. It is also difficult to extract robust distribution parameters from finite-size numerics \cite{schonle2021eigenstate, wang2022eigenstate}. The problem is especially severe in disordered systems, where matrix-element statistics are often broad and strongly non-Gaussian \cite{basko2006metal,luitz2016anomalous, serbyn2017thouless}.

Reduced dynamics of a subsystem coupled to a bath is often described by master (Lindblad) equations. This usually assumes Markovian dynamics with transition rates estimated by Fermi's golden rule (FGR). In strongly disordered interacting systems, both assumptions can fail because localization effects in Hilbert space become important \cite{anderson1958absence,altshuler1997quasiparticle,gornyi2005interacting,basko2006metal}. This phenomenon is known as many-body localization (MBL) \cite{nandkishore2015many, abanin2019colloquium}. Even outside the localized phase, broad distributions of relaxation rates are expected \cite{gornyi2005interacting,basko2006metal,tikhonov2021anderson}. As a result, the quantitative link between Hilbert-space transition rates and experimentally relevant real-space relaxation remains incomplete.

Recent quantum-processor experiments highlight this gap: real-space dynamics can appear frozen while relaxation in Hilbert space continues and follows broad, often power-law statistics \cite{lunkin2026evidence}. This behavior is naturally connected to spin-glass phenomenology \cite{sherrington1975solvable, thouless1977solution, mezard1987spin, georges2001quantum}. It also raises broader questions about slow modes, finite-size effects, and the delocalization crossover \cite{vsuntajs2020quantum, morningstar2022avalanches, sierant2022challenges, tikhonov2018many, doggen2018many, faoro2019non, lunkin2026evidence}. These developments motivate a dynamical framework that remains valid beyond the standard Markovian and FGR regimes.

In this work, we study a composite system consisting of a small thermometer and a large bath. We model interaction matrix elements in the non-interacting basis as independent random variables, without assuming finite variance. First, we derive a relation between reduced thermometer dynamics and the distribution of bath-state level broadenings. Second, we show that this relation yields thermalisation of the thermometer on a timescale set by the inverse typical broadening. We also derive a non-Markovian generalization of the global balance equation.

The structure of this work is as follows. In Sec.~\ref{sec: Diffusion propagator}, we introduce the setup and the main quantities. In Sec.~\ref{sec: cavity equation}, we derive the cavity equation, and in Sec.~\ref{sec: Diffuson} we generalize the approach to describe the statistics of off-diagonal Green-function matrix elements. In Sec.~\ref{sec: spin_1_2}, we test our predictions numerically. Finally, in Sec.~\ref{sec: discussion} we discuss the main implications and limitations of our results.
\section{Diffusion propagator.}
\label{sec: Diffusion propagator}
We consider a system described by the Hamiltonian
\begin{equation}
    \hat{H} = \hat{H}_{T} + \hat{H}_{B} + \hat{V}_{int},
\end{equation}
where $\hat{H}_{T}$ and $\hat{H}_{B}$ denote the Hamiltonians of the thermometer and the bath, respectively, and $\hat{V}_{int}$ describes their interaction. We use Latin and Greek indices to label eigenstates of the bath and the thermometer, respectively. Eigenstates of the non-interacting system ($\hat{V}_{int}=0$) are labeled by bold Latin indices $\pmb{j}$, or by pairs $(\alpha,j)$.

The joint probability that the thermometer is initially in state $\beta$ and is in state $\alpha$ at time $t>0$ is
\begin{align}
\label{eq: probability definition}
    p_{\alpha\beta}(t) = \tr\left[\hat{P}_{\alpha}(t) \hat{P}_{\beta} \hat{\varrho}\right].
\end{align}

Here $\hat{P}_{\alpha} \equiv |\alpha\rangle \langle \alpha|\otimes \hat{1}$ is the projector onto the thermometer state $|\alpha\rangle$, and $\hat{\varrho}$ denotes the density matrix of the full system.

We focus on the dynamics of diagonal elements of the reduced density matrix of the thermometer, assuming that the interaction does not renormalize the thermometer Hamiltonian. The dynamics of off-diagonal elements may play an important role in MBL and are left for future work.

We assume that the initial state has a well-defined energy; this is the only property of the density matrix relevant for our analysis. Under this assumption, the density matrix can be written as $\hat{\varrho}=f(\hat{H})$. The function $f$ is positive and satisfies the normalization condition $\int d\epsilon\,N\nu(\epsilon)f(\epsilon)=1$, where $\nu(\epsilon)$ is the normalized global density of states. We then introduce the distribution
\begin{equation}
    P(\epsilon) \equiv \frac{f(\epsilon)}{N\nu(\epsilon)}.
\end{equation}
The well-defined-energy condition means that this distribution is centered near a given energy and is narrower than the density of states. Then Eq.~(\ref{eq: probability definition}) can be rewritten as

\begin{align}
\label{eq: probability spectral}
    p_{\alpha\beta}(t)  = \int \frac{d \epsilon_+}{2\pi}\,\frac{d \epsilon_-}{2\pi} \,\tr\left[\hat{P}_{\alpha}\hat{G}\left(\epsilon_+ + i0\right)\hat{P}_{\beta} \left(\hat{G}\left(\epsilon_- - i0\right)  - \hat{G}\left(\epsilon_- + i0\right) \right)   \right]\frac{P(\epsilon_-)}{N\nu(\epsilon_-)}e^{-i(\epsilon_+ - \epsilon_-) t}.
\end{align}
Here we introduced the Green function of the full system, $\hat{G}(z)\equiv \left[z-\hat{H}\right]^{-1}$. At sufficiently long times, the leading contribution comes from terms whose Green-function poles lie in different half-planes. We therefore rewrite the expression as
\begin{equation}
\label{eq: probability simplified}
    p_{\alpha\beta}(t) \approx  \sum_{j,k}\int_{-\infty}^{\infty} \frac{d \epsilon}{2\pi} \int_{-i\infty}^{i\infty}\frac{d \varkappa}{2\pi i}  \mathcal{D}_{(\alpha, j),(\beta,k)}\frac{P(\epsilon)}{\nu(\epsilon)}e^{\varkappa t}.
\end{equation}
We define the propagator
\begin{align}
\label{eq: propagator definition}
    \mathcal{D}_{\pmb{j}\pmb{k}} = \tr\left[\hat{P}_{\pmb{j}}\hat{G}\left(\epsilon + i\frac{\varkappa}{2}\right)  \hat{P}_{\pmb{k}}\hat{G}\left(\epsilon - i\frac{\varkappa}{2}\right)\right],\quad \hat{P}_{(\alpha,j)}\equiv |\alpha\rangle \langle \alpha| \otimes |j\rangle\langle j|.
\end{align}
It is also convenient to introduce the propagator between thermometer states,
\begin{align}
\label{eq: averaged propagator}
  D_{\alpha,\beta}   = \frac{1}{N} \sum_{j,k} \mathcal{D}_{(\alpha, j),(\beta,k)} = \tr\left[\hat{P}_{\alpha}\hat{G}\left(\epsilon + i\frac{\varkappa}{2}\right)\hat{P}_{\beta}\hat{G}\left(\epsilon - i\frac{\varkappa}{2}\right)\right].
\end{align}
We show that this expression leads to thermalisation regardless of $P(\epsilon)$, provided that the distribution is sufficiently narrow and the interaction produces nonzero level broadening.

The sum over bath states can be interpreted as an average. To evaluate it, we assume that the matrix elements of $V_{\mathrm{int}}$ are independent random variables. This assumption can be viewed as an application of ETH to the interaction operator $V_{\mathrm{int}}$.

The propagator in Eq.~(\ref{eq: propagator definition}) involves products of Green-function matrix elements. We therefore start by analyzing diagonal matrix elements. This product of two Green functions is analytic in $\varkappa$ in the right half-plane. For convenience, we first take $\varkappa$ to be positive and real, and then perform analytic continuation.

\section{Cavity equation.}
\label{sec: cavity equation}

We treat the interaction term $V_{int}$ as a random matrix. This assumption simplifies the analysis in the large-matrix-size limit and leads to the cavity equation \cite{thouless1977solution}. The cavity approximation is exact when the corresponding hopping Hamiltonian describes motion on a tree graph. It is also applicable to Hamiltonians represented by large dense matrices \cite{cizeau1994theory, lunkin2025local}. The tree-graph analogy follows from a simple observation: for a random walk on the associated graph, the probability of forming a loop along a path of fixed length scales as $1/N$, where $N$ is the dimension of the Hamiltonian. In our approach, $N$ denotes the bath Hilbert-space dimension; this convention does not affect the estimate above. The cavity equation reads:
\begin{equation}
\label{eq: cavity}
\left(\hat{G}_{\pmb{j}\pmb{j}}\right)^{-1} = \left(\hat{G}^{(0)}_{\pmb{j}\pmb{j}}\right)^{-1}   - \Sigma_{\pmb{j}},\quad  \Sigma_{\pmb{j}} = \sum_{\pmb{k}} \left|V_{\pmb{j}\pmb{k}}\right|^2 \hat{G}^{\hat{\pmb{j}}}_{\pmb{k}\pmb{k}}.
\end{equation}
Here $\hat{G}^{(0)}$ denotes the Green function of the non-interacting system, and $\hat{G}^{\hat{\pmb{j}}}_{\pmb{k}\pmb{k}}$ is the Green function of the Hamiltonian with the $\pmb{j}$-th row and column removed. We use this notation for other cavity quantities below as well.

\begin{figure}
    \centering
    \includegraphics{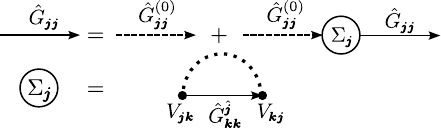}
    \caption{Diagrammatic representation of the cavity equation for the self-energy. Note the similarity to the standard non-crossing approximation, where the dashed line denotes a pairing according to Wick's theorem. Here, the crossed line indicates that we only consider terms where two perturbation-matrix elements are complex conjugates.}
    \label{fig: self-energy}
\end{figure}

The structure of the cavity equation resembles a diagrammatic self-energy calculation (see Fig.~\ref{fig: self-energy}). However, there is a qualitative difference between the two approaches. Standard crossed-diagram techniques assume a finite variance of matrix elements and therefore yield a self-averaging self-energy. In contrast, the cavity equation yields the full distribution of the self-energy, even when the second moment of matrix elements does not exist. If level broadening is self-averaging, it becomes independent of $\varkappa$, enabling the use of Fermi's golden rule. For broadly distributed matrix elements, the self-energy distribution depends on $\varkappa$ \cite{lunkin2025local}. This may imply the absence of a local-in-time equation of motion for thermalisation, so the Markovian approximation becomes inapplicable. Nevertheless, as we show below, a thermal distribution can still emerge in this regime.

\section{Emergence of diffusion}
\label{sec: Diffuson}

We extend the diagrammatic analogy introduced above to evaluate the propagator in Eq.~(\ref{eq: propagator definition}). As in the cavity-equation analysis, we begin with a Hamiltonian describing hopping on a tree. In this case, the off-diagonal Green-function matrix element between vertices $\pmb{j}_1$ and $\pmb{j}_n$ is
\begin{equation}
    \hat{G}_{\pmb{j}_1\pmb{j}_n} =  \hat{G}_{\pmb{j}_1\pmb{j}_1}\prod_{k = 1}^{n-1} H_{\pmb{j}_k \pmb{j}_{k+1}}\hat{G}^{\hat{\pmb{j}_k}}_{\pmb{j}_{k+1}\pmb{j}_{k+1}}.
\end{equation}
Here $\pmb{j}_1\ldots \pmb{j}_k \ldots \pmb{j}_n$ denotes the tree path connecting vertices $\pmb{j}_1$ and $\pmb{j}_n$.

We now switch to dense Hamiltonian matrices, where many paths connect a pair of vertices $\pmb{j}$ and $\pmb{k}$. In the large-$N$ limit, each path can be treated as effectively non-crossing, so we apply the tree expression to each path and sum over them. The propagator in Eq.~(\ref{eq: propagator definition}) contains a product of two Green functions, and we neglect interference terms in this product. This approximation is justified at times comparable to the inverse level spacing and leads to the following diffusive propagator:
\begin{equation}
\label{eq: diffuson paths}
    \mathcal{D}_{\pmb{j}\pmb{k}} = \sum_{n =0 }\sum_{\pmb{j_1}\ldots \pmb{j}_n| \pmb{j}_1 = \pmb{j}, \pmb{j}_n = \pmb{k}}\mathcal{B}_{\pmb{j}_1}\prod_{k = 1}^{n-1} |H_{\pmb{j}_k \pmb{j}_{k+1}}|^2 \mathcal{B}^{\hat{\pmb{j}_{k}}}_{\pmb{j}_{k+1}},\quad  \mathcal{B}^{\hat{\pmb{l}}}_{\pmb{j}} = \hat{G}^{\hat{\pmb{l}}}_{\pmb{j}\pmb{j}}\left(\epsilon + i \frac{\varkappa}{2}\right)\hat{G}^{\hat{\pmb{l}}}_{\pmb{j}\pmb{j}}\left(\epsilon - i \frac{\varkappa}{2}\right).
\end{equation}

\begin{figure}
    \centering
    \includegraphics{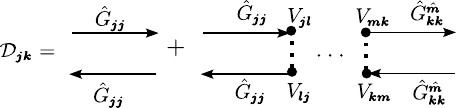}
    \caption{Diagrammatic representation of the leading contribution to the propagator from Eq.~(\ref{eq: propagator definition}). The argument of the top Green function is $\epsilon + i \varkappa/2$, whereas the argument of the bottom Green function is $\epsilon - i \varkappa/2$.}
    \label{fig: diffuson}
\end{figure}

This approximation is the analogue of ladder-diagram summation in standard diagrammatic techniques (see Fig.~\ref{fig: diffuson}).
In standard crossed-diagram techniques, the sum of such ladder diagrams is referred to as a diffuson, and the corresponding approximation is known as the diffusion approximation; we adopt the same terminology here. When $V_{\mathrm{int}}$ is treated as a random matrix with independent elements, this approximation gives the leading contribution to the propagator in $1/N$.

We introduce the following notation for further analysis. First, we define the local density of states $\nu$ and level broadening $\Gamma$:
\begin{align}
\label{eq: ldos}
    \tilde{\nu}_{\pmb{j}} \equiv  2\pi \nu_{\pmb{j}} \equiv  i\left[\hat{G}_{\pmb{j}\pmb{j}}\left(\epsilon + i \frac{\varkappa}{2}\right) - \hat{G}_{\pmb{j}\pmb{j}}\left(\epsilon - i \frac{\varkappa}{2}\right)\right], \nonumber\\
        \Gamma_{\pmb{j}} \equiv i\left[\Sigma_j\left(\epsilon + i \frac{\varkappa}{2}\right)  - \Sigma_j\left(\epsilon - i \frac{\varkappa}{2}\right)\right]=  \sum_{\pmb{k}}  \Gamma_{\pmb{j}\pmb{k}},\quad \Gamma_{\pmb{j}\pmb{k}} =  |V_{\pmb{j}\pmb{k}}|^2 \tilde{\nu}_{\pmb{k}}.
\end{align}
We rewrite the product of two Green functions in the form
\begin{equation}
\label{eq: product of G}
\mathcal{B}_{\pmb{j}} =\frac{\tilde{ \nu}_{\pmb{j}}}{\varkappa + \Gamma_{\pmb{j}}}.
\end{equation}
This correlator describes transitions in the full system. Each dashed line in Fig.~\ref{fig: diffuson} represents a transition between system states. Here we focus on transitions between thermometer states, so transition rates must be averaged over bath states. To evaluate the corresponding propagator, we sum over initial and final bath states according to Eq.~(\ref{eq: averaged propagator}).

In Eq.~(\ref{eq: diffuson paths}), the last factor $\mathcal{B}_{\pmb{j}_n}^{\hat{\pmb{j}}_{n-1}}$ in each product is only weakly correlated with previous factors. Correlations with other $\mathcal{B}$ terms are suppressed by $1/N$. Correlation with $|H_{\pmb{j}_{n-1},\pmb{j}_n}|^2$ is absent because this matrix element does not enter the definition of $\mathcal{B}_{\pmb{j}_n}^{\hat{\pmb{j}}_{n-1}}$. This allows us to replace this quantity by its average when evaluating the thermometer propagator in Eq.~(\ref{eq: averaged propagator}):
\begin{equation}
    \mathcal{B}_{\beta} = \frac{1}{N}\sum_{k} \mathcal{B}_{(\beta,k)}^{\hat{\pmb{j}}_{n-1}} \approx \frac{1}{N}\sum_{k} \mathcal{B}_{(\beta,k)}.
\end{equation}
However, to simplify the subsequent analysis, we need a slightly more involved replacement:
\begin{equation}
\mathcal{B}_{(\beta,k)}^{\hat{\pmb{j}}_{n-1}} \rightarrow \tilde{\nu}_{(\beta,k)}^{\hat{\pmb{j}}_{n-1}}\mathcal{B}_{\beta}/\tilde{\nu}_{\beta}, \quad
    \tilde{\nu}_{\beta} = \frac{1}{N}\sum_{k} \tilde{\nu}_{(\beta,k)}^{\hat{\pmb{j}}_{n-1}} \approx \frac{1}{N}\sum_{k} \tilde{\nu}_{(\beta,k)}.
\end{equation}
The same argument justifies this replacement: the correlation between $\tilde{\nu}_{(\beta,k)}^{\hat{\pmb{j}}_{n-1}}$ and all other terms is suppressed by $1/N$, so each factor can be replaced by its average, and vice versa.

We use this replacement to rewrite Eq.~(\ref{eq: averaged propagator}) as
\begin{equation}
\label{eq: diffuson averaged paths}
    \mathcal{D}_{\alpha \beta} = \left(\delta_{\alpha,\beta} +  \frac{1}{N} \sum_{j}\sum_{n = 1}\sum_{\pmb{j_1}\ldots \pmb{j}_n| \pmb{j}_1 = (\alpha,j)}\mathcal{B}_{\pmb{j}_1}\left[\prod_{k = 1}^{n-2} |H_{\pmb{j}_k \pmb{j}_{k+1}}|^2 \mathcal{B}^{\hat{\pmb{j}_{k}}}_{\pmb{j}_{k+1}}\right]\sum_{k} \frac{\Gamma_{\pmb{j}_{n-1},(\beta,k)}}{\tilde{\nu}_{\beta}}\right)\mathcal{B}_{\beta}.
\end{equation}
Next, we replace terms containing $\pmb{j}_{n-1}$ by their average over bath states. This replacement reads:
\begin{equation}
\label{eq: buble}
    \mathcal{B}^{\hat{\pmb{j}_{n-2}}}_{(\alpha_{n-1},j_{n-1})}\sum_{k} \frac{\Gamma_{\pmb{j}_{n-1},(\beta,k)}}{\tilde{\nu}_{\beta}} \rightarrow \hat{\Pi}_{\alpha_{n-1} , \beta} \equiv  \frac{1}{N\tilde{\nu}_{\beta}}\sum_{j,k} \mathcal{B}^{\hat{\pmb{j}_{n-2}}}_{(\alpha_{n-1},j_{n-1})}\Gamma_{(\alpha_{n-1},j_{n-1}),(\beta,k)}.
\end{equation}
Here we introduce the transition matrix $\Pi$. Repeating this analysis, we obtain the final expression for the propagator:
\begin{equation}
\label{eq: main}
    \mathcal{D}_{\alpha\beta} = \left(\sum_{n} \hat{\Pi}^n\right)_{\alpha\beta}\mathcal{B}_{\beta} = \left(\hat{1} - \hat{\Pi}\right)^{-1}_{\alpha\beta}\mathcal{B}_{\beta}.
\end{equation}

These equations summarize the central result of this work. They connect the real-space correlation function $\mathcal{D}$ to the level-broadening statistics $\Gamma$. This relation does not depend on the specific distribution of matrix elements of $V_{\mathrm{int}}$ or on the detailed form of the level-broadening distribution. For sufficiently large systems, the latter can be evaluated numerically using the cavity equation~(\ref{eq: cavity}), where this approach is computationally more efficient than exact diagonalization (see, e.g., Ref.~\cite{tikhonov2019critical}). The validity of Eq.~(\ref{eq: main}) reflects the applicability of the diffusion approximation to interacting disordered systems, which we test numerically in the simplest setting.

\section{Single spin thermalisation}
\label{sec: spin_1_2}
We consider a single-spin thermometer with level spacing $\Delta$ coupled to the bath. The Hamiltonian reads
\begin{equation}
    H = -\frac{\Delta}{2}\tau^z  + \tau^x \otimes V + H_{B}.
\end{equation}
Here $\tau^{z,x}$ are Pauli matrices.
In this case, only two types of transitions occur: from $0$ to $1$ and from $1$ to $0$. This allows the propagator to be simplified as
\begin{equation}
   \Pi_{\alpha\beta}  =  (\tilde{\nu}_{\alpha} - \varkappa \,\mathcal{B}_{\alpha})/\tilde{\nu}_{\beta}.
\end{equation}

Using this identity, the propagator becomes
\begin{equation}
\label{eq: main spin}
    \hat{\mathcal{D}} = \frac{ \tilde{\nu}_0 \tilde{\nu}_1 \left(\begin{smallmatrix} \mathcal{B}_0 & \frac{\mathcal{B}_1}{ \tilde{\nu}_1}\left(\tilde{\nu}_0 - \mathcal{B}_0 \varkappa\right) \\  \frac{\mathcal{B}_0}{\tilde{\nu}_0}\left(\tilde{\nu}_1 - \mathcal{B}_1 \varkappa\right) & \mathcal{B}_1\end{smallmatrix}\right)}{\varkappa \left( \tilde{\nu}_1\mathcal{B}_0 + \tilde{\nu}_0\mathcal{B}_1 - \mathcal{B}_0\mathcal{B}_1 \varkappa\right)}.
\end{equation}

We connect this propagator to the $\tau^z$ autocorrelation function using Eq.~(\ref{eq: probability definition}). The result reads
\begin{align}
   \tr \left[\tau^z(t) \tau^z \rho \right] = \int \frac{d\epsilon}{2\pi}\frac{d \varkappa}{2\pi i}  A(\epsilon,\varkappa) e^{\varkappa t} \frac{e^{-\beta \epsilon}}{Z}, \\
   A(\epsilon,\varkappa) \equiv \left(\begin{smallmatrix}
      1 \\ -1 
   \end{smallmatrix}\right)^T \hat{\mathcal{D}} \left(\begin{smallmatrix}
      1 \\ -1 
   \end{smallmatrix}\right).
   \label{eq: spin corr}
\end{align}
Here $Z = \int d\epsilon \nu(\epsilon)e^{-\beta \epsilon}$ denotes the partition function.
We use this relation to test our theoretical prediction. We compute $A(\epsilon,\varkappa)$ from Eq.~(\ref{eq: averaged propagator}) and compare it with the prediction of Eq.~(\ref{eq: main spin}) for several models. At sufficiently long times, the correlation function is dominated by its disconnected part, i.e., by the product of two average magnetizations at the given temperature. As a result, $A$ has a pole at $\varkappa = 0$. We therefore focus on models with parity symmetry, where the average spin is zero and this pole is absent, which improves precision at small frequencies.

We consider several models to verify our prediction. We start with a model in which $H_B$ is, by construction, a random matrix, and then we consider models of interacting spins.

\subsection{L{\'e}vy model.}
\label{subsec: Levy model}
\begin{figure}[H]
    \centering
    \includegraphics{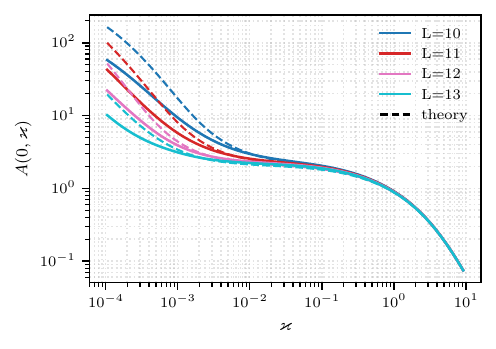}
    \caption{The Laplace transform of the $zz$ autocorrelation function for the L{\'e}vy model is plotted for different bath sizes $N \equiv 2^L$. The solid lines represent the direct evaluation of Eq.~(\ref{eq: averaged propagator}). The dashed lines represent the theoretical evaluation of Eq.~(\ref{eq: main spin}). We use the parameters $\mu = 1.5$, $\Delta = 0.9$, and $g = 0.4$. All data are averaged over at least 30 realizations.}
    \label{fig: Levy}
\end{figure}

In the first model, $H_{B}$ is a diagonal matrix with elements drawn from a normal distribution with unit width. The matrix elements of $V$ are i.i.d. random variables. Their signs are random, while their absolute values follow a Pareto distribution:
\begin{equation}
    P(|v|) = \frac{g^{\mu}}{N}\frac{\mu}{|v|^{1+\mu}}.
\end{equation}
Matrices with a power-law distribution of elements have been studied in Refs.~\cite{cizeau1994theory, tarquini2016level} and are known as L{\'e}vy matrices; we therefore refer to this ensemble as the L{\'e}vy model. The matrix-element scaling with $N$ is chosen so that the level broadening remains finite in the limit $N\rightarrow \infty$, enabling meaningful comparisons between matrices of different sizes. We use this ensemble to test Eq.~(\ref{eq: main}) in the absence of a finite second moment of interaction matrix elements. The model is further motivated by Ref.~\cite{long2023phenomenology}, where power-law matrix-element statistics emerge during the Jacobi algorithm. We compute both $D_{\alpha,\beta}$ and $\mathcal{B}_{\alpha}$ using exact diagonalization. The corresponding numerical results are shown in Fig.~\ref{fig: Levy}.

Curves for the same system size diverge at low frequencies. This difference originates from interference effects, which become important on the level-spacing scale $\delta_N \sim 1/N$. Another finite-size effect arises from the finite number of $\Gamma$ values used to compute the averages in Eq.~(\ref{eq: main}), and it is more pronounced for heavy-tailed distributions of $\Gamma$. This also explains the differences between solid curves for different system sizes. As the system size increases, both sets of curves converge to the same limiting behavior.

\subsection{TFIM}
\label{subsec: TFIM}
As a second example, we consider a transverse-field Ising model (TFIM) with all-to-all interactions:
\begin{equation}
    H_B = -\frac{1}{2}\sum_{j,k}J_{jk}\sigma^x_j\sigma^x_k + \sum_j h \sigma_j^z, \quad  V = \frac{g}{\sqrt{L}}\sum_{j = 1}^{L} \sigma_j^x.
\end{equation}
Here, the coupling constants $J_{jk}$ are drawn from a Gaussian distribution with width $1/\sqrt{L}$, where $L$ is the number of spins in the bath and $h = 1$. 

\begin{figure}[H]
    \centering
    \includegraphics{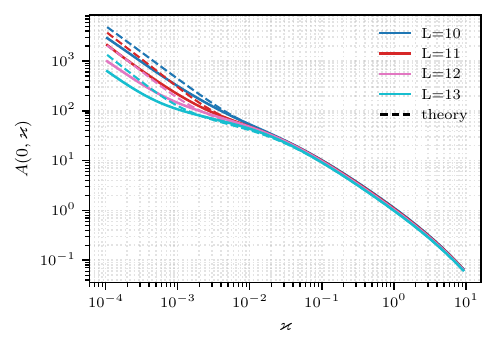}
    \caption{The Laplace transform of the $zz$ autocorrelation function for the TFIM is plotted for different bath sizes $L$. The solid lines represent the direct evaluation of Eq.~(\ref{eq: averaged propagator}). The dashed lines represent the theoretical evaluation of Eq.~(\ref{eq: main spin}). We use the parameters $\Delta = 0.9$ and $g = 0.4$. All data are averaged over at least 30 realizations.}
    \label{fig: TFIM}
\end{figure}

This model provides an example of a chaotic spin system, at least for states with energies near the center of the spectrum, where ETH is expected to hold. In contrast to the L{\'e}vy model, the number of random parameters scales only logarithmically with the Hilbert-space dimension. The interaction scaling is chosen to yield a finite scaling limit in the $L \rightarrow \infty$ regime. The corresponding numerical results are shown in Fig.~\ref{fig: TFIM}.

\subsection{Imbrie model.}
\label{subsec: Imbrie} 
\begin{figure}[H]
    \centering
    \includegraphics{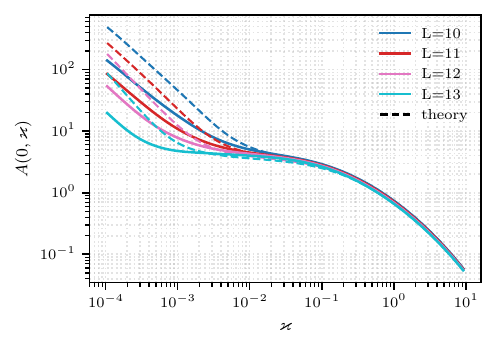}
    \caption{The Laplace transform of the $zz$ autocorrelation function for the Imbrie model is plotted for different bath sizes $L$. The solid lines represent the direct evaluation of Eq.~(\ref{eq: averaged propagator}). The dashed lines represent the theoretical evaluation of Eq.~(\ref{eq: main spin}). We use the parameters $\Delta = 0.9$, $g = 0.4$, and $W = 2$. All data are averaged over at least 30 realizations.}
    \label{fig: Imbrie}
\end{figure}

As a final example, the bath is described by the model commonly called the Imbrie model \cite{biroli2024large, imbrie2016many}. The Hamiltonian of this one-dimensional model has the form
\begin{equation}
H_B = \sum_j \Delta_j \sigma_j^z \sigma_{j+1}^z + \sum_j \left( h_j \sigma_j^z + \Gamma  \sigma_j^x\right),\quad V = g \sigma_0^x.
\end{equation}
Here $\Gamma = 1$, while $\Delta_i$ and $h_i$ are i.i.d. random variables drawn uniformly from $[0.8,1.2]$ and $[-W/2,W/2]$, respectively. We use this model as an example of a one-dimensional interacting system with parity symmetry. The system localizes for sufficiently large disorder $W$. The numerical results are shown in Fig.~\ref{fig: Imbrie} for $W = 2$. For larger $W$, the Hamiltonian eigenfunctions become more localized, which leads to poorer self-averaging of $\hat{\Pi}$. Equivalently, at sufficiently large disorder the spins are almost independent, so the thermometer interacts effectively only with spin $0$ and the diffusive picture is no longer relevant. As a result, agreement worsens at higher $W$.

Overall, the agreement between the theoretical evaluation based on the diffusion picture and exact calculations is reasonable for all three models.

\section{Discussion and conclusion}
\label{sec: discussion}

The structure of Eq.~(\ref{eq: main}) implies a conserved-probability mode. At $\varkappa=0$, the matrix $\hat{1}-\hat{\Pi}$ has a zero eigenvalue with right eigenvector $\tilde{\nu}_{\beta}$, which generates the zero-frequency pole of the propagator. The existence of this pole is a manifestation of probability conservation. This pole controls the long-time-limit behavior of transition probabilities:
\begin{equation}
    p^{(\infty)}_{\alpha\beta} = \int d\epsilon\, \frac{\nu_{\alpha}(\epsilon)}{\nu(\epsilon)}\, s_{\beta}(\epsilon)\, \mathcal{B}_{\beta}(\epsilon,\varkappa{=}0)\, P(\epsilon),
\end{equation}
where $s_{\beta}$ is the left zero mode of $\hat{1}-\hat{\Pi}$, and $\nu(\epsilon)=\sum_{\alpha}\nu_{\alpha}(\epsilon)$. Let us introduce the bath density of states $\nu_B(\epsilon)$. If it is sufficiently broad, we can relate it to $\nu_{\alpha}(\epsilon) = \nu_B(\epsilon - \xi_{\alpha})$. We also assume that the distribution $P(\epsilon)$ is centered near the energy $E$ and that its width is negligible compared to the width of $\nu_B(\epsilon)$. Using this notation, we can introduce the effective temperature:

\begin{equation}
    \beta(E) \equiv \frac{d}{dE}\ln\nu_B(E),
\end{equation}
The narrow-distribution assumption allows us to write the probabilities in the following form:
\begin{align}
    p^{(\infty)}_{\alpha\beta} &= p^{\mathrm{Th}}_{\alpha}\, \bigl[s_{\beta}\mathcal{B}_{\beta}\bigr]_{\epsilon=E,\varkappa=0}, \nonumber\\
    p^{\mathrm{Th}}_{\alpha} &\equiv \frac{\nu_B(E-\xi_{\alpha})}{\sum_{\gamma}\nu_B(E-\xi_{\gamma})}
    = \frac{e^{-\beta(E)\xi_{\alpha}}}{\sum_{\gamma}e^{-\beta(E)\xi_{\gamma}}}.
    \label{eq: Gibbs}
\end{align}
This formula reflects that the system thermalises. In particular, stationary probabilities should also be thermal, i.e., $p^{(\infty)}_{\alpha\beta}=p^{\mathrm{Th}}_{\alpha}p^{\mathrm{Th}}_{\beta}$. This implies
\begin{equation}
    s_{\beta} \propto \frac{\tilde{\nu}_{\beta}}{\mathcal{B}_{\beta}}\Big|_{\varkappa=0}.
\end{equation}
This relation generalizes the global-balance condition beyond Markovian kinetics. We consider the following example to illustrate it. In the self-averaging regime (e.g., Gaussian-distributed matrix elements of $V_{\mathrm{int}}$), one can define effective rates
\begin{equation}
    \Gamma_{\alpha \rightarrow \beta} \equiv \sum_k \Gamma_{(\alpha,j),(\beta,k)}, \qquad \Gamma_{\alpha} \equiv \sum_{\beta} \Gamma_{\alpha \rightarrow \beta}.
\end{equation}
Then $\mathcal{B}_{\alpha}=\tilde{\nu}_{\alpha}/(\varkappa+\Gamma_{\alpha})$, and the zero-mode condition gives
\begin{equation}
    \sum_{\alpha}\left(\Gamma_{\beta\rightarrow\alpha}\tilde{\nu}_{\beta}-\tilde{\nu}_{\alpha}\Gamma_{\alpha\rightarrow\beta}\right)=0,
\end{equation}
which is the familiar global-balance equation consistent with FGR and detailed balance.

An important caveat is the regime where the typical broadening vanishes, $\Gamma_{\mathrm{typ}}\to 0$ as $\varkappa\to 0$. In that case, the stationary distribution need not be thermal. The L{\'e}vy model with $\mu<1$ provides a representative example \cite{cizeau1994theory}. This regime is closely related to MBL physics \cite{altshuler1997quasiparticle, gornyi2005interacting, basko2006metal}, and our results complement that literature by linking real-space relaxation to level-broadening statistics in Hilbert space.

In summary, we derived a diffusion-approximation expression for the reduced dynamics of a small thermometer coupled to a many-body bath [Eq.~(\ref{eq: main})], identified its relation to level-broadening statistics, and verified the prediction numerically in three distinct bath models (L{\'e}vy, all-to-all TFIM, and the Imbrie model; Figs.~\ref{fig: Levy}, \ref{fig: TFIM}, \ref{fig: Imbrie}).

Several directions are natural next steps. First, extending the formalism to off-diagonal elements of the reduced density matrix would clarify dephasing in the same framework. Second, applying the approach to transport observables may help characterize slow modes in disordered interacting systems \cite{capizzi2025hydrodynamics}. Finally, the diffusion approximation may offer practical simplifications for numerical simulations of chaotic many-body dynamics \cite{haghshenas2025digital, mandra2025heuristic}.

\section*{Acknowledgements}
The author gratefully acknowledges K. Tikhonov and I. Gorny for insightful discussions and valuable comments on the manuscript. The author also thanks M. Feigel’man for helpful suggestions.

\bibliography{bibliography.bib}

@article{srednicki1994chaos,
  title={Chaos and quantum thermalization},
  author={Srednicki, Mark},
  journal={Physical review e},
  volume={50},
  number={2},
  pages={888},
  year={1994},
  publisher={APS}
}

@article{deutsch1991quantum,
  title={Quantum statistical mechanics in a closed system},
  author={Deutsch, Josh M},
  journal={Physical review a},
  volume={43},
  number={4},
  pages={2046},
  year={1991},
  publisher={APS}
}

@article{srednicki1999approach,
  title={The approach to thermal equilibrium in quantized chaotic systems},
  author={Srednicki, Mark},
  journal={Journal of Physics A: Mathematical and General},
  volume={32},
  number={7},
  pages={1163},
  year={1999},
  publisher={IOP Publishing}
}

@article{d2016quantum,
  title={From quantum chaos and eigenstate thermalization to statistical mechanics and thermodynamics},
  author={D'Alessio, Luca and Kafri, Yariv and Polkovnikov, Anatoli and Rigol, Marcos},
  journal={Advances in Physics},
  volume={65},
  number={3},
  pages={239--362},
  year={2016},
  publisher={Taylor \& Francis}
}

@article{deutsch2018eigenstate,
  title={Eigenstate thermalization hypothesis},
  author={Deutsch, Joshua M},
  journal={Reports on Progress in Physics},
  volume={81},
  number={8},
  pages={082001},
  year={2018},
  publisher={IOP Publishing}
}

@article{schonle2021eigenstate,
  title={Eigenstate thermalization hypothesis through the lens of autocorrelation functions},
  author={Sch{\"o}nle, Christoph and Jansen, David and Heidrich-Meisner, Fabian and Vidmar, Lev},
  journal={Physical Review B},
  volume={103},
  number={23},
  pages={235137},
  year={2021},
  publisher={APS}
}

@article{wang2022eigenstate,
  title={Eigenstate thermalization hypothesis and its deviations from random-matrix theory beyond the thermalization time},
  author={Wang, Jiaozi and Lamann, Mats H and Richter, Jonas and Steinigeweg, Robin and Dymarsky, Anatoly and Gemmer, Jochen},
  journal={Physical Review Letters},
  volume={128},
  number={18},
  pages={180601},
  year={2022},
  publisher={APS}
}

@article{thouless1977solution,
  title={Solution of'solvable model of a spin glass'},
  author={Thouless, David J and Anderson, Philip W and Palmer, Robert G},
  journal={Philosophical Magazine},
  volume={35},
  number={3},
  pages={593--601},
  year={1977},
  publisher={Taylor \& Francis}
}

@article{cizeau1994theory,
  title={{Theory of L{\'e}vy matrices}},
  author={Cizeau, Pierre and Bouchaud, Jean-Philippe},
  journal={Physical Review E},
  volume={50},
  number={3},
  pages={1810},
  year={1994},
  publisher={APS}
}

@article{lunkin2025local,
  title={{Local density of states correlations in the L{\'e}vy-Rosenzweig-Porter random matrix ensemble}},
  author={Lunkin, Aleksey V and Tikhonov, Konstantin},
  journal={SciPost Physics},
  volume={19},
  number={1},
  pages={015},
  year={2025}
}

@article{abanin2019colloquium,
  title={Colloquium: Many-body localization, thermalization, and entanglement},
  author={Abanin, Dmitry A and Altman, Ehud and Bloch, Immanuel and Serbyn, Maksym},
  journal={Reviews of Modern Physics},
  volume={91},
  number={2},
  pages={021001},
  year={2019},
  publisher={APS}
}

@article{gornyi2005interacting,
  title={Interacting electrons in disordered wires: Anderson localization and low-T transport},
  author={Gornyi, Igor V and Mirlin, Alexander D and Polyakov, Dmitry G},
  journal={Physical review letters},
  volume={95},
  number={20},
  pages={206603},
  year={2005},
  publisher={APS}
}

@article{basko2006metal,
  title={Metal--insulator transition in a weakly interacting many-electron system with localized single-particle states},
  author={Basko, Denis M and Aleiner, Igor L and Altshuler, Boris L},
  journal={Annals of physics},
  volume={321},
  number={5},
  pages={1126--1205},
  year={2006},
  publisher={Elsevier}
}

@article{altshuler1997quasiparticle,
  title={Quasiparticle lifetime in a finite system: A nonperturbative approach},
  author={Altshuler, Boris L and Gefen, Yuval and Kamenev, Alex and Levitov, Leonid S},
  journal={Physical review letters},
  volume={78},
  number={14},
  pages={2803},
  year={1997},
  publisher={APS}
}

@article{haghshenas2025digital,
  title={Digital quantum magnetism at the frontier of classical simulations},
  author={Haghshenas, Reza and Chertkov, Eli and Mills, Michael and Kadow, Wilhelm and Lin, Sheng-Hsuan and Chen, Yi-Hsiang and Cade, Chris and Niesen, Ido and Begu{\v{s}}i{\'c}, Tomislav and Rudolph, Manuel S and others},
  journal={arXiv preprint arXiv:2503.20870},
  year={2025}
}

@article{mandra2025heuristic,
  title={A Heuristic for Matrix Product State Simulation of Out-of-Equilibrium Dynamics of Two-Dimensional Transverse-Field Ising Models},
  author={Mandr{\`a}, Salvatore and Astrakhantsev, Nikita and Isakov, Sergei and Villalonga, Benjamin and Ware, Brayden and Westerhout, Tom and Kechedzhi, Kostyantyn},
  journal={arXiv preprint arXiv:2511.23438},
  year={2025}
}

@article{sherrington1975solvable,
  title={Solvable model of a spin-glass},
  author={Sherrington, David and Kirkpatrick, Scott},
  journal={Physical review letters},
  volume={35},
  number={26},
  pages={1792},
  year={1975},
  publisher={APS}
}

@book{mezard1987spin,
  title={Spin glass theory and beyond: An Introduction to the Replica Method and Its Applications},
  author={M{\'e}zard, Marc and Parisi, Giorgio and Virasoro, Miguel Angel},
  volume={9},
  year={1987},
  publisher={World Scientific Publishing Company}
}

@article{georges2001quantum,
  title={Quantum fluctuations of a nearly critical Heisenberg spin glass},
  author={Georges, Antoine and Parcollet, Olivier and Sachdev, Subir},
  journal={Physical Review B},
  volume={63},
  number={13},
  pages={134406},
  year={2001},
  publisher={APS}
}

@article{capizzi2025hydrodynamics,
  title={Hydrodynamics and the eigenstate thermalization hypothesis},
  author={Capizzi, Luca and Wang, Jiaozi and Xu, Xiansong and Mazza, Leonardo and Poletti, Dario},
  journal={Physical Review X},
  volume={15},
  number={1},
  pages={011059},
  year={2025},
  publisher={APS}
}

@article{serbyn2017thouless,
  title={Thouless energy and multifractality across the many-body localization transition},
  author={Serbyn, Maksym and Papi{\'c}, Z and Abanin, Dmitry A},
  journal={Physical Review B},
  volume={96},
  number={10},
  pages={104201},
  year={2017},
  publisher={APS}
}

@article{luitz2016anomalous,
  title={Anomalous thermalization in ergodic systems},
  author={Luitz, David J and Bar Lev, Yevgeny},
  journal={Physical review letters},
  volume={117},
  number={17},
  pages={170404},
  year={2016},
  publisher={APS}
}

@article{tarquini2016level,
  title={Level statistics and localization transitions of L{\'e}vy matrices},
  author={Tarquini, Elena and Biroli, Giulio and Tarzia, Marco},
  journal={Physical review letters},
  volume={116},
  number={1},
  pages={010601},
  year={2016},
  publisher={APS}
}

@article{anderson1958absence,
  title={Absence of diffusion in certain random lattices},
  author={Anderson, Philip W and others},
  journal={Physical review},
  volume={109},
  number={5},
  pages={1492--1505},
  year={1958}
}

@article{tikhonov2021anderson,
  title={From Anderson localization on random regular graphs to many-body localization},
  author={Tikhonov, Konstantin S and Mirlin, Alexander D},
  journal={Annals of Physics},
  volume={435},
  pages={168525},
  year={2021},
  publisher={Elsevier}
}

@article{nandkishore2015many,
  title={Many-body localization and thermalization in quantum statistical mechanics},
  author={Nandkishore, Rahul and Huse, David A},
  journal={Annu. Rev. Condens. Matter Phys.},
  volume={6},
  number={1},
  pages={15--38},
  year={2015},
  publisher={Annual Reviews}
}

@article{faoro2019non,
  title={Non-ergodic extended phase of the quantum random energy model},
  author={Faoro, Lara and Feigel’man, Mikhail V and Ioffe, Lev},
  journal={Annals of Physics},
  volume={409},
  pages={167916},
  year={2019},
  publisher={Elsevier}
}

@article{lunkin2026evidence,
  title={Evidence for a two-dimensional quantum glass state at high temperatures},
  author={Lunkin, Aleksey and Ticea, Nicole S and Kumar, Shashwat and Miao, Connie and Choi, Jaehong and Alghadeer, Mohammed and Drozdov, Ilya and Abanin, Dmitry and Abbas, Amira and Acharya, Rajeev and others},
  journal={arXiv preprint arXiv:2601.01309},
  year={2026}
}

@article{vsuntajs2020quantum,
  title={Quantum chaos challenges many-body localization},
  author={{\v{S}}untajs, Jan and Bon{\v{c}}a, Janez and Prosen, Toma{\v{z}} and Vidmar, Lev},
  journal={Physical Review E},
  volume={102},
  number={6},
  pages={062144},
  year={2020},
  publisher={APS}
}

@article{morningstar2022avalanches,
  title={Avalanches and many-body resonances in many-body localized systems},
  author={Morningstar, Alan and Colmenarez, Luis and Khemani, Vedika and Luitz, David J and Huse, David A},
  journal={Physical Review B},
  volume={105},
  number={17},
  pages={174205},
  year={2022},
  publisher={APS}
}

@article{sierant2022challenges,
  title={Challenges to observation of many-body localization},
  author={Sierant, Piotr and Zakrzewski, Jakub},
  journal={Physical Review B},
  volume={105},
  number={22},
  pages={224203},
  year={2022},
  publisher={APS}
}

@article{doggen2018many,
  title={Many-body localization and delocalization in large quantum chains},
  author={Doggen, Elmer VH and Schindler, Frank and Tikhonov, Konstantin S and Mirlin, Alexander D and Neupert, Titus and Polyakov, Dmitry G and Gornyi, Igor V},
  journal={arXiv preprint arXiv:1807.05051},
  year={2018}
}

@article{tikhonov2018many,
  title={Many-body localization transition with power-law interactions: Statistics of eigenstates},
  author={Tikhonov, KS and Mirlin, AD},
  journal={Physical Review B},
  volume={97},
  number={21},
  pages={214205},
  year={2018},
  publisher={APS}
}

@article{long2023phenomenology,
  title={Phenomenology of the prethermal many-body localized regime},
  author={Long, David M and Crowley, Philip JD and Khemani, Vedika and Chandran, Anushya},
  journal={Physical Review Letters},
  volume={131},
  number={10},
  pages={106301},
  year={2023},
  publisher={APS}
}

@article{tikhonov2019critical,
  title={Critical behavior at the localization transition on random regular graphs},
  author={Tikhonov, KS and Mirlin, AD},
  journal={Physical Review B},
  volume={99},
  number={21},
  pages={214202},
  year={2019},
  publisher={APS}
}

@article{biroli2024large,
  title={Large-deviation analysis of rare resonances for the many-body localization transition},
  author={Biroli, Giulio and Hartmann, Alexander K and Tarzia, Marco},
  journal={Physical Review B},
  volume={110},
  number={1},
  pages={014205},
  year={2024},
  publisher={APS}
}

@article{imbrie2016many,
  title={On many-body localization for quantum spin chains},
  author={Imbrie, John Z},
  journal={Journal of Statistical Physics},
  volume={163},
  number={5},
  pages={998--1048},
  year={2016},
  publisher={Springer}
}




\nolinenumbers

\end{document}